\documentclass{PoS}
\usepackage{amsmath}
\usepackage{cite}
\usepackage{textcomp}
\title{Recent developments in nonleptonic kaon decays}

\ShortTitle{Recent developments in nonleptonic kaon decays}

\author{\speaker{Oscar Cat\`a}\thanks{This work was supported in part by the DFG cluster of excellence 'Origin and Structure of the Universe' and was performed in the context of the ERC Advanced Grant project 'FLAVOUR' (267104).}\\
        Ludwig-Maximilians-Universit\"at M\"unchen, Fakult\"at f\"ur Physik,
Arnold Sommerfeld Center for Theoretical Physics, 
D--80333 M\"unchen, Germany
\\
        E-mail: \email{oscar.cata@physik.uni-muenchen.de}}

\abstract{
I review the current status of nonleptonic kaon decays, placing special emphasis on the recent theoretical progress. In particular, I concentrate on 3 points: (i) the improved determination of $\epsilon_K$, including both perturbative and nonperturbative contributions; (ii) the efforts to tame $K\to 2\pi$ transitions in lattice QCD; and (iii) the use of holographic methods to solve the vector meson dominance puzzle in $K\to 3\pi$.}

\FullConference{2013 Kaon Physics International Conference,\\
		29 April-1 May 2013\\
		University of Michigan, Ann Arbor, Michigan - USA}

\begin{document}
%%%%%%%%%%%%%%%%%%%%%%%%%%%%%%%%%%%%%%%%%%%%%%%%%%%%%%%%%%%%%%%%%%%%%%%%%%%%%
\section{A short primer in kaon CP violation}
CP violation in kaon physics is a mature and well-established field in particle physics. Far from being exhaustive, the present Section is only meant to introduce the basic concepts and set the notations that I will use later on. For details, I refer the reader to the many excellent reviews that exist in the literature ({\it{e.g.}}, \cite{deRafael:1995zv,D'Ambrosio:1996nm,Cirigliano:2011ny}).  
 
Weak interactions induce mixing between the strong eigenstates $K^0$ and ${\bar{K^0}}$, which are related to the CP eigenstates $K_1$ and $K_2$ as
\begin{align}
K_{1,2}=\frac{1}{\sqrt{2}}(K^0\mp {\bar{K^0}})
\end{align}
This explains why $K_S$ (which is mostly $K_1$) decays into $2\pi$ while $K_L$ (mostly $K_2$) into $3\pi$. However, the fact that $K_L\to 2\pi$ is nonzero points at CP violation, which is regulated by the small parameter ${\tilde{\epsilon}}$:
\begin{align}
K_L=\frac{1}{\sqrt{1+|{\tilde{\epsilon}}|^2}}(K_2+\tilde{\epsilon}K_1)
\end{align}
In order to study CP violation, it is convenient to work with the $K\to 2\pi$ amplitudes in the isospin decomposition and define the ratios
\begin{align}
\epsilon_K=\frac{A(K_L\to (\pi\pi)_0)}{A(K_S\to (\pi\pi)_0)};\quad \omega=\frac{A(K_S\to (\pi\pi)_2)}{A(K_S\to (\pi\pi)_0)};\quad \chi=\frac{A(K_L\to (\pi\pi)_2)}{A(K_S\to (\pi\pi)_0)}
\end{align}
The previous ratios describe both direct and indirect CP violation. Indirect CP violation is described by $\epsilon_K$, while direct CP violation is accessible through the combination $\displaystyle\epsilon^{\prime}_K\equiv\frac{\chi-\epsilon_K\cdot \omega}{\sqrt{2}}$. 

At the experimental level, one has access to the amplitudes
\begin{align}
\eta_{+-}=\frac{A(K_L\to \pi^+\pi^-)}{A(K_S\to \pi^+\pi^-)};\qquad
\eta_{00}=\frac{A(K_L\to \pi^0\pi^0)}{A(K_S\to \pi^0\pi^0)}
\end{align}
in terms of which $\epsilon_K$ and $\epsilon^{\prime}_K$ can be determined as 
\begin{align}
|\epsilon_K|=\frac{1}{3}\Big(2|\eta_{+-}|+|\eta_{00}|\Big);\qquad 
{\mathrm{Re}}\left(\frac{\epsilon^{\prime}}{\epsilon}\right)_K&=\frac{1}{3}\left(1-\left|\frac{\eta_{00}}{\eta_{+-}}\right|\right)
\end{align}
Indirect CP violation was confirmed experimentally in 1964, while direct CP violation was only experimentally established after the KTeV and NA48 measurements (see Figure~1). 
\begin{figure}[h]
\begin{center}
\includegraphics[width=8.5cm]{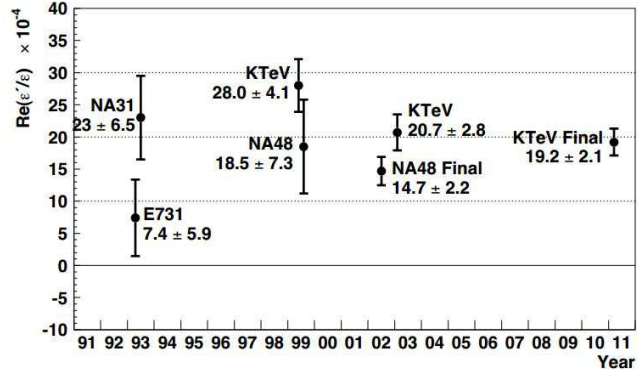}
\end{center}
\vskip -0.7cm
\caption{\small{\it{Evolution of the determination of $\epsilon^{\prime}$ with time. Figure taken from Ref.~\cite{Worcester}.}}}\label{fig:1}
\end{figure}
The current world average values are~\cite{Beringer}
\begin{align}
|\eta_{00}|&=2.220(11)\cdot 10^{-3}\qquad 
|\eta_{+-}|=2.232(11)\cdot 10^{-3}\nonumber\\
|\epsilon_K|&=2.228(11)\cdot 10^{-3}\qquad \quad\,\,\,\phi_{\epsilon}=(43.52\pm0.05)^{\circ}\nonumber\\
{\mathrm{Re}}\left(\frac{\epsilon^{\prime}}{\epsilon}\right)_K&=1.66(23)\cdot 10^{-3}\qquad \quad\,\,\,\,\,\, \phi_{\epsilon^{\prime}}=(42.3\pm1.5)^{\circ}
\end{align}
From a theoretical point of view, the determination of $\epsilon_K$ and $\epsilon^{\prime}_K$ requires a dynamical description of the $K^0-{\bar{K^0}}$ and $K\to 2\pi$ matrix elements. These processes are described by the effective $\Delta S=1,2$ Hamiltonians~\cite{Gilman:1979bc,Gilman:1982ap,Buchalla:1995vs} 
\begin{align}
H_{eff}^{\Delta S=1}&=\frac{G_F}{\sqrt{2}}\lambda_u\sum_{i=1}^{10}\Big[z_i(\mu)+\tau y_i(\mu)\Big]Q^{S=1}_i(\mu)\nonumber\\
H_{eff}^{\Delta S=2}&=\frac{G_F^2}{4\pi^2}\Big[\lambda_c^2F_1+\lambda_t^2F_2+2\lambda_c\lambda_tF_3\Big]Q^{S=2}(\mu)
\end{align}
At the relevant energies, {\it{i.e.}} close to the kaon masses, they can be mapped into ChPT operators:
\begin{align}
H_{eff}^{\Delta S=1}&=\frac{G_F}{\sqrt{2}}f_0^4\lambda_u\left\{g_8{\cal{O}}_8+g_{27}{\cal{O}}_{27}+e^2g_{ew}{\cal{O}}_{ew}\right\}+ {\rm{NLO}}\nonumber\\
H_{eff}^{\Delta S=2}&=-\frac{G_F^2}{16\pi^2}f_0^4g_{2}{\cal{O}}_{2}+{\rm{NLO}}
\end{align}
where ($L_{\mu}=UD_{\mu}U^{\dagger}$)
\begin{align}
{\cal{O}}_8&=\langle L_{\mu}L^{\mu}\rangle_{23};\quad {\cal{O}}_{27}=\langle L_{\mu}\rangle_{23}\langle L^{\mu}\rangle_{11}+\frac{2}{3}\langle L_{\mu}\rangle_{21}\langle L^{\mu}\rangle_{13};\quad {\cal{O}}_2=\langle L_{\mu}\lambda_{23}L^{\mu}\lambda_{23}\rangle
\end{align}
The connection with $\epsilon_K$ can be worked out from the neutral kaon mixing matrix element
\begin{align}
M_{21}=\frac{1}{2m_K}\left[\langle {\bar{K}}^0|H_{eff}^{\Delta S=2}|K^0\rangle+\sum_{n}\frac{\langle {\bar{K}}^0|H_{eff}^{\Delta S=1}|n\rangle\langle n|H_{eff}^{\Delta S=1}|K^0\rangle}{m_K-E_n+i\epsilon}\right]
\end{align}
whose real part is connected to $\Delta m_K$, while the imaginary part is proportional to $\epsilon_K$. Defining
\begin{align}
A(K^0\to (\pi\pi)_I)=A_Ie^{i\delta_I};\qquad A({\bar{K}}^0\to (\pi\pi)_I)=A_I^*e^{i\delta_I} 
\end{align}
and assuming that the intermediate states in $(\Delta S=1)^2$ are dominated by $\pi\pi$ exchange, one finds that  
\begin{align}
\epsilon_K&=e^{i\phi_{\epsilon}}\sin\phi_{\epsilon}\left[\frac{{\rm{Im}}\langle{\bar{K}}^0|H_{eff}^{\Delta S=2}|K^0\rangle}{\Delta m_K}+\frac{{\rm{Im}}A_0}{{\rm{Re}}A_0}\right]
\end{align}
Using the previous result, $\epsilon^{\prime}$ can be expressed, to a very good approximation, as 
\begin{align}
\epsilon^{\prime}_K&=\frac{i}{\sqrt{2}}e^{i(\delta_2-\delta_0)}\frac{{\mathrm{Re}}A_2}{{\mathrm{Re}}A_0}\left[\frac{{\mathrm{Im}}A_2}{{\mathrm{Re}}A_2}-\frac{{\mathrm{Im}}A_0}{{\mathrm{Re}}A_0}\right]
\end{align}

%%%%%%%%%%%%%%%%%%%%%%%%%%%%%%%%%%%%%%%%%%%%%%%%%%%%%%%%%%%%%%%%%%%%%%%%%%%%%

\section{$\Delta S=2$ transitions: $B_K$ and $\epsilon_K$}
Indirect CP violation in $K\to 2\pi$ is responsible for the $K^0-{\bar{K^0}}$ mixing. The contributions to $\epsilon_K$ are given by $\Delta S=2$ box diagrams and can be written down as 
\begin{align}\label{definition}
\epsilon_K&=e^{i\phi_{\epsilon}}\sin\phi_{\epsilon}\left[\frac{{\rm{Im}}\langle{\bar{K}}^0|H_{eff}^{\Delta S=2}|K^0\rangle}{\Delta m_K}+\frac{{\rm{Im}}A_0}{{\rm{Re}}A_0}\right]
\end{align}
The first term is the local $\Delta S=2$ transition, which consists of a short and long-distance contribution, and is given by the effective Hamiltonian~\cite{Buchalla:1995vs}
\begin{align}
H_{eff}^{\Delta S=2}&=\frac{G_F^2m_W^2}{16\pi^2}\Big[\lambda_c^2S_0(x_c)\eta_1+\lambda_t^2S_0(x_t)\eta_2+2\lambda_c\lambda_tS_0(x_c,x_t)\eta_3\Big]c_i(\mu)Q^{\Delta S=2}(\mu)
\end{align}
The term in brackets collects the Inami-Lim functions with electroweak and strong perturbative corrections. Long distances are described by a single (multiplicatively renormalizable) operator $Q^{\Delta S=2}=({\bar{s}}_L\gamma_{\mu}d_L)({\bar{s}}_L\gamma^{\mu}d_L)$, whose matrix element in $K^0-{\bar{K^0}}$ mixing defines the so-called $B_K$ bag parameter, which is a genuine nonperturbative object. For convenience, it is common to work with the RG-invariant ${\hat{B}}_K$:
\begin{align}
\langle K^0|c_i(\mu) Q^{\Delta S=2}(\mu)|{\bar{K^0}}\rangle &=\frac{8}{3}f_K^2m_K^2{\hat{B}}_K
\end{align}

The second part in Eq.~(\ref{definition}) is a purely long-distance $(\Delta S=1)^2$ piece, which gives a subleading (but nonnegligible) contribution to $\epsilon_K$. In the last years, there has been progress on both the perturbative and nonperturbative contributions. Regarding ${\hat{B}}_K$, at present the best determinations come from lattice simulations~\cite{Colangelo:2010et}
\begin{align}\label{lat}
\hat{B}_{K}\Big|_{N_f=2+1} = 0.738(20);\qquad\qquad\qquad \hat{B}_{K}\Big|_{N_f=2} = 0.729(25)(17)
\end{align}
Determinations with analytical methods cannot compete with the lattice precision but are nonetheless essential to understand the previous numbers. In particular, combining the chiral and large-$N_c$ expansions has proven to be very effective. At leading order in both expansions one finds $B_K=0.75$, which substantially improves the vacuum saturation approximation, $B_K^{VS}=1$. When $1/N_c$ corrections are included one is sensitive to the scale-dependence and a careful matching between long and short distances has to be done~\cite{Bardeen:1987vg}. ${\cal{O}}(1/N_c)$ corrections turn out to be sizable and negative, but they are compensated to a large extent by sizable and positive chiral ${\cal{O}}(p^4)$ contributions. As a result, the final number barely changes to $B_K=0.70(10)$~\cite{Bardeen:1987vg}. Comparison with the lattice results shows remarkable agreement. 

However, the situation is not entirely satisfactory. In the chiral limit, $B_K$ is known from the relation~\cite{Donoghue:1982cq} 
\begin{align}\label{chi}
B_K^{\chi}&=\frac{5}{4}g_{27}\sim 0.37
\end{align}
which holds to all orders in the momentum expansion. Using the $1/N_c$ expansion with proper long and short-distance matching, the previous result was successfully reproduced already at ${\cal{O}}(p^4,1/N_c)$ in a series of works by different groups~\cite{Bijnens:1995br,Peris:2000sw,Cata:2003mn,Bijnens:2006mr}. However, comparison between Eq.~(\ref{lat}) and Eq.~(\ref{chi}) indicates that mass corrections should bring a huge contribution, accounting for roughly 50\% of the value of $B_K$. Attempts to compute the mass corrections consistently within chiral perturbation theory have fallen systematically short of the lattice value. As far as I know, the issue of mass corrections in $B_K$ has not yet been fully understood.
  
One of the most interesting implications of the current lattice precision on $B_K$ is that one can no longer dismiss the $(\Delta S=1)^2$ nonperturbative contributions to $\epsilon_K$~\cite{Buras:2008nn}. These extra long-distance effects can be parametrized in terms of an overall prefactor $\kappa_{\epsilon}$ as follows
\begin{align}\label{param}
\epsilon_K&=e^{i\phi_{\epsilon}}\sin\phi_{\epsilon}\left[\frac{{\mathrm{Im}}M_{12}}{\Delta m_K}+\rho\frac{{\mathrm{Im}}A_0}{{\mathrm{Re}}A_0}\right]\equiv \kappa_{\epsilon}\frac{e^{i\phi_{\epsilon}}}{\sqrt{2}}\left[\frac{{\mathrm{Im}}M_{12}}{\Delta m_K}\right]
\end{align}
If $\rho=1$, one obtains the estimate $\kappa_{\epsilon}\sim 0.92(2)$~\cite{Buras:2008nn}. However, as initially observed in~\cite{Bijnens:1990mz}, the $(\Delta S=1)^2$ contribution is related by a dispersion relation to nonlocal (other than $B_K$) long distances in $M_{12}$, with some cancellation between both effects.\footnote{Local dimension-eight contributions to $M_{12}$ can be shown to be negligible~\cite{Cata:2004ti}.} Taking both effects into account~\cite{Buras:2010pza} $\rho$ gets reduced to $\rho=0.6(3)$ and accordingly 
\begin{align}\label{res}
\kappa_{\epsilon}&=0.94(2)
\end{align}
In Eq.~(\ref{param}), it is implicitly understood that the first term corresponds to the $B_K$ contribution, while $\rho$ collects the non-$B_K$ contributions. Using the values of $\kappa_{\epsilon}$ and $B_K$ on the nonperturbative side and the perturbative corrections to NNLO~\cite{Brod:2010mj}, the latest theoretical result for $\epsilon_K$ reads~\cite{Brod:2011ty}
\begin{align}
|\epsilon_K|=1.90(26)\cdot 10^{-3}
\end{align}
which falls a bit shy of the experimental number.

As noticed in~\cite{Buras:2008nn}, failure to fit the experimental value for $\epsilon_K$ leads to some tension in the CKM fit between the K and B systems. Specifically, using the parameterization
\begin{align}
|\epsilon_K|\sim \kappa_{\epsilon}f_K^2{\hat{B}}_K|V_{cb}|^4\xi_s^2\frac{C_s}{C_d}\sin 2\beta
\end{align}
the suppression induced by $\kappa_{\epsilon}$ combined with lower values of $B_K$ would require a slightly too large $\sin 2\beta$, which would conflict with $B_s$ data. A way out would be to invoke new CP-violating phases, {\it{e.g.}} $S_{\psi K_s}=\sin(2\beta+2\phi_d)$. If this tension is of eventual significance remains to be seen. What seems to be on a rather good handle is the value of ${\kappa_{\epsilon}}$. Lattice simulations for the absorptive part find~\cite{Blum:2012uk}
\begin{align}
(\kappa_{\epsilon})_{\rm{abs}}=0.924(6)
\end{align}
which is in excellent agreement with the analytical estimate reported in~\cite{Buras:2008nn}. 

%%%%%%%%%%%%%%%%%%%%%%%%%%%%%%%%%%%%%%%%%%%%%%%%%%%%%%%%%%%%%%%%%%%%%%%%%%%%%

\section{Recent progress in $\Delta S=1$ transitions}
The determination of $\epsilon^{\prime}_K$ boils down to an understanding of the so-called $\Delta I=1/2$ rule and the contributions of the $Q_6\sim {\mathrm{Im}}\, A_0$ and $Q_8\sim {\mathrm{Im}}\, A_2$ matrix elements. Their separate influence on $\epsilon^{\prime}_K$ can be seen below:
\begin{align}
\epsilon^{\prime}_K&=\frac{i}{\sqrt{2}}e^{i(\delta_2-\delta_0)}\frac{{\mathrm{Re}}A_2}{{\mathrm{Re}}A_0}\left[\frac{{\mathrm{Im}}A_2}{{\mathrm{Re}}A_2}-\frac{{\mathrm{Im}}A_0}{{\mathrm{Re}}A_0}\right]
\end{align}
The main difficulty from a theoretical point of view is to understand the $\Delta I=1/2$ rule puzzle, namely why ${\mathrm{Re}}A_0\simeq 22.5\, {\mathrm{Re}}A_2$ is roughly 15 times bigger than expected from naive factorization. 

While at present there is no solid quantitative understanding of the $\Delta I=1/2$ rule, at least there is widespread consensus on the following qualitative points: 
\begin{itemize}
\item The RG-mixing of the current-current operators as they evolve down in energy can account for roughly 10\% of the enhacement. 
\item The bulk of the corrections come from nonperturbative effects, where enhanced hadronic matrix elements should bring in 90\% of the effect.
\item Penguin contributions get enhanced at hadronic energies and are an important ingredient to explain the size of $A_0$. 
\end{itemize}
Considerable quantitative progress has been achieved by combining different nonperturbative methods with the large-$N_c$ expansion~\cite{Bardeen:1986vp,Bardeen:1986vz,Bijnens:2000im,Hambye:1999yy,Hambye:2003cy}. For instance, it has been realized that non-factorizable contributions are sizeable and point in the right direction. However, quantitative improvement on the determination of hadronic matrix elements is extremely challenging. As it happened with $B_K$, lattice QCD simulations can be an extremely useful tool here. However, $K\to 2\pi$ decays are much more challenging to simulate than $K^0-{\bar{K^0}}$ mixing and that has hindered progress on $\Delta S=1$ transitions for a long time. This situation might however have reached a tipping point. Quite recently, in a series of papers appearing in the last 2 years~\cite{Blum:2012uk, Boyle:2012ys}, the RBC-UKQCD collaboration has released results that hint at a solution of the $\Delta I=1/2$ rule puzzle. Specifically, they reported~\cite{Boyle:2012ys} an accidental cancellation of contributions (let me abstractly denote them as $t_1$ and $t_2$) in $A_2$. This cancellation is closely linked to the breakdown of factorization: instead of $t_2\simeq t_1/3$ they obtain $t_2\simeq -0.7 t_1$. Interestingly, the same contributions appear in $A_0$ but with different signs, such that no cancellation is observed there. Schematically, the overall picture that emerges is
\begin{align}
\frac{{\mathrm{Re}}A_0}{{\mathrm{Re}}A_2}\sim \frac{2t_2-t_1}{t_1+t_2}
\end{align}
There are a number of reasons to be optimistic about this result. First of all, $A_2$ has been simulated down to the physical masses, giving~\cite{Blum:2012uk}
\begin{align}
{\rm{Re}}A_2=1.381(46)(258)\cdot 10^{-8}\,{\rm{GeV}};\qquad {\rm{Im}}A_2=-6.54(46)(120)\cdot 10^{-13}\,{\rm{GeV}}
\end{align}
These results have to be refined to reduce the systematic errors for a more meaningful comparison, but so far they are in good agreement with experiment, ${\rm{Re}}A_2=1.479(4)\cdot 10^{-8}\,{\rm{GeV}}$. Second of all, while $A_0$ is more challenging and up to now simulations are still at unphysical masses, naive extrapolation of what they have observed so far gets in the ballpark of the experimental value. This naive extrapolation has to be taken with a grain of salt, but it should be a good indication, especially given the mild mass dependence observed in $A_0$. Finally, their results confirm the smallness of penguins at perturbative scales. For more details, see the talks by Norman Christ and Robert Mawhinney at this conference. So far the lattice results are in good overall agreement with the qualitative features pointed out in previous large-$N_c$-based studies~\cite{Bardeen:1986vp,Bardeen:1986vz}, which is certainly reassuring. However, to complete the picture, it would be very interesting if the lattice could assess how much of an enhancement penguins get at low energies and their actual impact on $A_0$. 

%%%%%%%%%%%%%%%%%%%%%%%%%%%%%%%%%%%%%%%%%%%%%%%%%%%%%%%%%%%%%%%%%%%%%%%%%%%%%

\section{Experimental and theoretical status of $K\to 3\pi$}
CP conserving $K\to 3\pi$ modes admit the general decomposition~\cite{Devlin:1978ye}
\begin{eqnarray}\label{k3pi}
{\cal{M}}(K_L\to \pi^+\pi^-\pi^0)&=&\alpha_1-\beta_1u+(\zeta_1+\xi_1)u^2+\frac{1}{3}(\zeta_1-\xi_1)v^2~,\nonumber\\
{\cal{M}}(K_L\to \pi^0\pi^0\pi^0)&=&-3\alpha_1-\zeta_1(3u^2+v^2)~,\nonumber\\
{\cal{M}}(K^+\to \pi^+\pi^+\pi^-)&=&2\alpha_1+\beta_1u+(2\zeta_1-\xi_1)u^2+\frac{1}{3}(2\zeta_1+\xi_1)v^2~,\nonumber\\
{\cal{M}}(K^+\to \pi^+\pi^0\pi^0)&=&-\alpha_1+\beta_1u-(\zeta_1+\xi_1)u^2-\frac{1}{3}(\zeta_1-\xi_1)v^2
\end{eqnarray}
where I have kept only the dominant octet contributions. $u,v,s_i$ and $s_0$ are kinematic variables
\begin{equation}
u=\frac{s_3-s_0}{m_{\pi}^2}~,\qquad v=\frac{s_1-s_2}{m_{\pi}^2}~,\qquad  s_i=(p_K-p_{\pi_i})^2~,\qquad s_0=\frac{1}{3}\sum_{i=1}^3 s_i
\end{equation} 
while $\alpha_1,\beta_1,\zeta_1,\xi_1$ are dynamical parameters that can be expressed in terms of the low-energy couplings of the chiral (strong and electroweak) Lagrangian. At NLO one finds~\cite{Kambor:1991ah}
\begin{eqnarray}
\alpha_1&=&\alpha_1^{(0)}-\frac{2g_8}{27f_Kf_{\pi}}m_K^4\left\{(k_1-k_2)+24{\cal{L}}_1\right\}~,\nonumber\\
\beta_1&=&\beta_1^{(0)}-\frac{g_8}{9f_Kf_{\pi}}m_{\pi}^2m_K^2\left\{(k_3-2k_1)-24{\cal{L}}_2\right\}~,\nonumber\\
\zeta_1&=&-\frac{g_8}{6f_Kf_{\pi}}m_{\pi}^4\left\{k_2-24{\cal{L}}_1\right\}~,\nonumber\\
\xi_1&=&-\frac{g_8}{6f_Kf_{\pi}}m_{\pi}^4\left\{k_3-24{\cal{L}}_2\right\}
\end{eqnarray} 
where ${\cal{L}}_i$ collect the strong low-energy couplings and $k_i$ the electroweak ones. The structure of the counterterms makes it manifest that $K\to 3\pi$ processes involve strong amplitudes with weak external vertices as well as direct weak terms. 
Specifically, one finds
\begin{align}
{\cal{L}}_1&={\cal{L}}_2+3L_2=2L_1+2L_2+L_3\nonumber\\
k_1&=9(-N_5+2N_7-2N_8-N_9)\nonumber\\
k_2&=3(N_1+N_2+2N_3)\nonumber\\
k_3&=3(N_1+N_2-N_3)
\end{align}
The previous results were recomputed in~\cite{Bijnens:2002vr} and fits to data were made including isospin and electromagnetic corrections. A good overall phenomenological fit to data was found~\cite{Bijnens:2004ai}. From a theoretical viewpoint, however, one would like to understand the dynamics behind the values of the low-energy couplings. That goes beyond the scope of ChPT and one has to adopt some hadronic-scale models. In the strong sector, vector meson dominance has proven to be a more than acceptable mechanism to estimate the low-energy couplings~\cite{Ecker:1988te}. The same idea was exported to the electroweak sector in the so-called factorization models~\cite{Isidori:1991ya,Ecker:1992de,D'Ambrosio:1997tb}, where resonance exchange was assumed to dominate both the strong and electroweak low-energy couplings. This leads to relations between $N_i$ and $L_i$ and enhances the predictive power in the electroweak sector. Still, due to the large number of weak counterterms, the accuracy of the hadronic models is hard to test. When applied to $K\to 3\pi$, all the factorization models found a set of generic features:
\begin{itemize}
\item[(i)] $k_1$ is dominated by the scalar meson sector.
\item[(ii)] $k_2,k_3$ are affected mostly by vector meson exchange and related to the strong counterterms as follows:
\begin{align}
&k_2=24{\cal{L}}_1=0~,\nonumber\\
&k_3=24\left({\cal{L}}_2+\frac{3}{4}L_9\right)=24\left(L_3+\frac{3}{4}L_9\right)
\end{align}
where the second equality can be linked to the Skyrme structure of the ${\cal{O}}(p^4)$ strong ChPT Lagrangian.
\item[(iii)] Strong cancellations between the strong and weak diagrams are to be expected.
\end{itemize}
Using the results from vector meson dominance in the strong sector~\cite{Ecker:1988te}, one is led to conclude that $k_3=0$, which violates the vector meson dominance hypothesis and, worst of all, is in contradiction with experimental data~\cite{:2008js,Batley:2010fj}, which seems to favor instead $k_3\sim 5\cdot 10^{-9}$. The fact that there are strong cancellations (see last point above) already indicates that $k_3=0$ might be a fine-tuned solution instead of a generic result. However, the absence of a model with $k_3\neq 0$ was definitely puzzling.

In Ref.~\cite{Cappiello:2011re} a model for the electroweak chiral Lagrangian was introduced based on the gauge/gravity duality~\cite{Maldacena:1997re,Gubser:1998bc,Witten:1998qj}. In these settings, the Standard Model fields live in a 4-dimensional boundary brane, while a fifth dimension is responsible for the strong interactions, conjectured to be dual to a weakly-coupled gravitational theory with Anti-de Sitter (AdS) geometry. The boundaries can thus be seen as probes of the strong interactions. In~\cite{Cappiello:2011re} it was shown that introducing the electroweak interactions as double-trace perturbations in the boundary~\cite{Witten:2001ua} is equivalent to a factorization model for the electroweak interactions, where both the strong and electroweak low-energy couplings are determined in terms of the AdS geometry of the 5-dimensional bulk space. Remarkably, in that model $L_3=-\frac{11}{24}L_9$ and $k_3\sim 3\cdot 10^{-9}$, showing that, contrary to~\cite{Isidori:1991ya,Ecker:1992de,D'Ambrosio:1997tb}, compliance with experiment can be achieved within vector meson dominance. The seeming failure of vector meson dominance in $K\to 3\pi$ was therefore not generic but a model-dependent artifact.

I will conclude this Section with some brief comments on the status of CP violation in $K\to 3\pi$ decays. Here, for once, theory is ahead of experiment. Regarding the indirect CP violation, KLOE has recently improved the bound on $K_S\to 3\pi^0$ to~\cite{Babusci:2013tr} 
\begin{align}
{\rm{Br}}(K_S\to 3\pi^0)< 2.8\cdot 10^{-8}
\end{align}
which is still one order of magnitude above the Standard Model estimate at $1.9\cdot 10^{-9}$. For more details, see the talk of Patrizia De Simone in this conference.

For direct CP violation, NA48/2 has values for the slope asymmetries compatible with no signal at the $10^{-4}$ level~\cite{Batley:2007aa}, while the Standard Model expectation is at $10^{-5}$~\cite{Gamiz:2003pi}.

%%%%%%%%%%%%%%%%%%%%%%%%%%%%%%%%%%%%%%%%%%%%%%%%%%%%%%%%%%%%%%%%%%%%%%%%%%%%%

\section{Summary and future directions}

Kaon physics has a rather mature status and a long track of experimental successes. Indirect and direct CP violation are nowadays known, respectively, within a 5\textperthousand~ and 14\% accuracy, $K^0-{\bar{K^0}}$ mixing still being the most stringent flavor test for new physics models. However, there are still some long-standing fundamental issues that remain unexplained. In this paper I have mentioned the tension between the theoretical prediction of $\epsilon_K$ and its experimental number, which persists after nonperturbative effects and NNLO perturbative corrections are accounted for. On the direct CP side, a quantitative understanding of the $\Delta I=1/2$ rule is still pending, despite the efforts of the community over the years. On the experimental side, it is not yet settled whether the amount of CP violation in $K\to 3\pi$ fits the Standard Model prediction. 

Improvements on each of those aspects are hard to achieve and might be perceived from an outsider's perspective as slow, but they are steady. An example is the promising path recently opened in lattice QCD to determine $\epsilon^{\prime}_K$, which is making solid headway and will provide, in the coming years, a most wanted determination of ${\rm{Re}}(\epsilon^{\prime}/\epsilon)_K$. $\Delta m_K$ is also in the agenda. A clean determination of the short-distance {\it{vs}} long-distance budget in this quantity would be a valuable tool to constraint new-physics scenarios. 

It is hard to overstate the importance of such determinations. However, it would certainly be unsatisfactory to consider them a solution without supplementing them with a deeper analytical understanding than the one we have today. The recent lattice progress should thus also serve as both stimulus and guidance to continue improving on the theoretical analytical side.  

%%%%%%%%%%%%%%%%%%%%%%%%%%%%%%%%%%%%%%%%%%%%%%%%%%%%%%%%%%%%%%%%%%%%%%%%%%%%%

\end{document}